\documentclass[prl,twocolumn,aps,showpacs]{revtex4}%
\usepackage{graphicx}
\usepackage{amsfonts}
\usepackage{bm}
\usepackage{amsmath}
\usepackage{amssymb}%
\begin{document}
\title{Efficient Magnetization Reversal with Noisy Currents }
\author{Wouter Wetzels, Gerrit E. W. Bauer and Oleg N. Jouravlev}
\affiliation{Kavli Institute of NanoScience, Delft University of Technology, Lorentzweg 1,
2628 CJ Delft, The Netherlands}


\begin{abstract}
We propose to accelerate reversal of the ferromagnetic order parameter in spin
valves by electronic noise. By solving the stochastic equations of motion we
show that the current-induced magnetization switching time is drastically
reduced by a modest level of externally generated current (voltage) noise.
This also leads to a significantly lower power consumption for the switching process.

\end{abstract}
\maketitle

The dynamics of the ferromagnetic order parameter persists to pose a
challenging problem of fundamental and applied nature \cite{general}. With
increasing bit density of mass data storage devices and\ emergence of the
magnetic random access memory (MRAM) concept, the speed and energy dissipation
of the magnetization switching process have become important issues.
In the present MRAM generation, magnetic bits are written by spatially extended
\O rsted magnetic fields, which sets limits to bit size and power consumption. An attractive alternative
method is the current-induced magnetization switching predicted by
theoreticians \cite{slonczewski96,berger96} and confirmed experimentally in
nano-pillar devices \cite{albert00,katine00,grollier01}. In these spin valves,
which consist basically of an electrically connected ferromagnet%
$\vert$%
normal%
$\vert$%
ferromagnetic metals sandwich, the electric current is polarized in a
\textquotedblleft fixed\textquotedblright\ layer of high magnetic coercivity
and exerts a \textquotedblleft spin-transfer torque\textquotedblright\ on the
second magnetically soft ferromagnet as sketched in Fig. 1. Recently,
time-resolved measurements of the current-induced magnetization dynamics have
been reported \cite{krivorotov05}. Advanced theoretical models for the
spin and charge transport in magnetic devices
\cite{waintal00,brataas00,stiles02} lead to a reasonable description of the
magnetization dynamics within the macro-spin model, in which the magnetization
is assumed to move rigidly under external magnetic field and spin-transfer
torques \cite{apalkov04,manschot04}. We should note that in larger devices
evidence has been found for spin waves and more complicated excitations that
require full micromagnetic simulations \cite{lee04}.

The main obstacle that prevents wide application of current-induced switching
is the high critical current needed to reverse the magnetization. There 
are proposals on how to reduce the critical
current by sample design \cite{manschot04} and optimizing the switching
process by a precessional switching strategy \cite{kent04}. Since the
spin-transfer torque vanishes for the collinear stable point of a spin valve,
the switching time depends strongly on processes that induce a canting between
the magnetizations such that the spin-transfer torque starts to kick in. This
happens for example by increasing temperature, and is the basic idea of the
pre-charging strategy by Devolder \textit{et al}. \cite{Devolder}. More
advanced strategies used in conventional magnetization switching require
pulse-shaped microwaves \cite{sun05} and rely on precise knowledge of the
magnetization dynamics with proper feedback.

Electrical and magnetization noise usually degrades device and system
performance and often efforts have to be undertaken to reduce it as much as
possible. In nonlinear systems intentionally added noise may e.g. enhance the
quality of signal transmission by the phenomenon of stochastic resonance.
Noise generators find useful application, for example to test the response of
a system to noise, to generate random signals for use in in encryption or to
minimize the effect of quantization errors by a method called dithering. In
this Letter, we propose a simple method to improve the energy-efficiency of
current-induced magnetization switching by adding noise to the electric
circuit connected to the device. We demonstrate that this leads to a reversal
process with increased switching speed and less energy dissipation.

\begin{figure}[tbh]
\centering
\includegraphics[width=0.85\columnwidth]{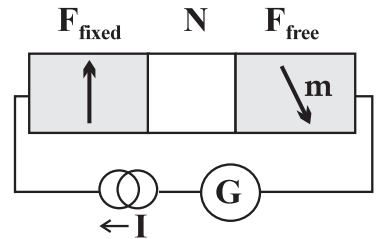}\caption{Schematic picture
of the spin valve under consideration. The applied potential on the left/right
side is $\pm V/2$. The charge current $I$ flows clockwise and G is a
noise generator.}%
\label{schematic}%
\end{figure}

The magnetization dynamics in the macro-spin model for a small magnetic grain
is described by the Landau-Lifshitz-Gilbert (LLG) equation \cite{brown63}
\begin{equation}
\dot{\bm m}=-\gamma\mathbf{m}\times\bm H_{\mathrm{eff}}+\alpha\mathbf{m}%
\times\dot{\mathbf{m}} \label{LLG}%
\end{equation}
where $\gamma$ is the gyromagnetic ratio and $\alpha$ the Gilbert damping
constant. The magnetization direction $\mathbf{m}=\left(
\cos\varphi\sin\theta,\sin\varphi\sin\theta,\cos\theta\right)  $ is
parameterized by the angles $\theta$ and $\varphi,\ $and we consider switching
between $\theta=0$ (magnetization pointing up) and $\theta=\pi$ (down). When a 
current bias is applied to the device in Fig. \ref{schematic}%
, the LLG equation for the free layer $F_{\mathrm{free}}$ needs to be modified
to include the spin-transfer torque $\dot{\bm m}_{\mathrm{torque}}%
=-(\gamma/{\cal M}_{S})\mathbf{m}\times\mathbf{I}_{\mathrm{free}}%
\times\mathbf{m}$, where ${\cal M}_{S}$ is the total magnetic moment of the ferromagnet. 
The electric charge and spin currents driven through the systems fluctuate due to thermal and shot noise
\cite{foros05}. Here we wish to investigate the effect of intentionally
applied current fluctuations with spectral density and bandwidth controlled by
an external noise generator. The fluctuating (spin) current through the
interfaces of $F_{\mathrm{free}}$ creates a fluctuating spin torque that can
be taken into account by adding a fluctuating torque to the LLG equation,
$\dot{\bm m}_{\mathrm{fluct}}\equiv\mathbf{m}\times\left(  \mathbf{h}%
\times\mathbf{m}\right)  ,$ where we introduced an effective random field
$\mathbf{h}=h_{\theta}\widehat{\theta}+h_{\phi}\widehat{\phi}+h_{r}\widehat
{r}$ in spherical coordinates, with $\left\langle h_{i}\left(  t\right)
\right\rangle =0\ $and $\left\langle h_{i}\left(  t\right)  h_{j}\left(
t^{\prime}\right)  \right\rangle =\mu_{ij}\delta_{ij}\left(  t-t^{\prime
}\right)  $ for\ $i,j\in\left\{  \theta,\phi,r\right\}  $. Such a model befits an
electromagnetic environment with a large number of degrees of freedom that
generates Gaussian noise with correlation times much shorter than the
magnetization response time. At room temperature, shot noise does not have to
be taken into account \cite{foros05}. The LLG equation in the presence of
random magnetic fields leads to a stochastic equation for the three components
of the unit vector $\bm m,$ which can be reduced to the Fokker-Planck equation
\cite{brown63}. This allows us to access the magnetization reversal time as a
function of applied switching current and current noise.

The switchable ferromagnet is assumed to have a uniaxial shape anisotropy,
with easy axis along $z,$ i.e. $\mathbf{H}_{\mathrm{eff}}=H_{K}\cos
\theta\widehat{z}$. For thermal stability of the equilibrium magnetization the ratio of the height of
the effective energy barrier and the thermal energy should be high. For our
model, this means that $H_{K}{\cal M}_{S}\gg$ $k_{B}T$ .

We use the formalism developed in \cite{brataas00} to find the charge and spin
currents through the spin valve. The ferromagnet-normal metal contacts are
assumed to have identical spin-dependent conductances $g^{\uparrow\uparrow}$
and $g^{\downarrow\downarrow}$ (in units of the conductance quantum $e^{2}/h$)
with $p=\left(  g^{\uparrow\uparrow}-g^{\downarrow\downarrow}\right)  /\left(
g^{\uparrow\uparrow}+g^{\downarrow\downarrow}\right)  $. We approximate the
mixing conductance $g^{\uparrow\downarrow},$ the interface-specific parameter
for the spin-transfer torque, by $\left(  g^{\uparrow\uparrow}+g^{\downarrow
\downarrow}\right)  /2$, which is sufficiently accurate for our purposes. The
spin accumulation in the normal metal in quasi-equilibrium can be found by
using current conservation and spin conservation in the normal metal.
Spin-flip in the normal metal is disregarded, while that in the ferromagnet is
absorbed in the spin-dependent conductances. Under an applied voltage bias
$V$, we then find a charge current
\begin{equation}
\frac{I}{eV}=\frac{e}{2h}\left(  g^{\uparrow\uparrow}+g^{\downarrow\downarrow
}\right)  \left(  1-p^{2}\sin^{2}\frac{\theta}{2}\right)  ,\label{ChCurrent}%
\end{equation}
and a spin current leaving the free ferromagnetic layer for the normal metal
node%
\begin{equation}
\frac{\mathbf{I}_{\mathrm{free}}}{eV}=\frac{g^{\uparrow\uparrow}%
-g^{\downarrow\downarrow}}{16\pi}\left(  -2\mathbf{m}\cos^{2}\frac{\theta}%
{2}+\widehat{\mathbf{\theta}}\sin\theta\right)  \,.\label{SpCurrent}%
\end{equation}
By inserting this expression into $\dot{\bm m}_{\text{torque}}$, we find for the spin torque contribution to the
dynamics.
\begin{equation}%
\begin{array}
[c]{ccc}%
\dot{\bm m}_{\mathrm{torque}} & = & \gamma J_{S}\sin\theta\,\widehat
{\mathbf{\theta}},\\
J_{S} & = & \left(  g^{\uparrow\uparrow}-g^{\downarrow\downarrow}\right)
eV/(16\pi {\cal M}_{S}).
\end{array}
\end{equation}
Since this torque vanishes at $\theta=0,\pi$, the absence of
(temperature-induced) fluctuations would imply an infinite switching time.
Combining all contributions discussed above, and disregarding an
angle-dependence of the Gilbert damping due to spin pumping
\cite{tserkovnyak02}, the LLG equation reduces to
\begin{align}
\dot{\theta} &  =-\gamma\alpha H_{K}\cos\theta\sin\theta-\gamma J_{S}%
\sin\theta+h_{\theta}\;,\nonumber\\
\dot{\varphi} &  =\gamma H_{K}\cos\theta+h_{\varphi}/\sin\theta\;.\label{StEq}%
\end{align}

In the following we focus on the effect of current fluctuations $\Delta
I\left(  t\right)  =I\left(  t\right)  -\left\langle I\right\rangle $ with
spectrum%
\begin{equation}
S\left(  \omega\right)  =\int dte^{i\omega t}\left\langle \Delta
I\left(  t\right)  \Delta I\left(  0\right)  \right\rangle ,\nonumber
\end{equation}
that are generated externally on top of the thermal noise. The noise spectral
density is taken to be a constant $S_{w}$ over a bandwidth $\Delta\omega$ that
depends on the typical time scales of the system: the system is not sensitive
to fluctuations slower than the timescale of the magnetization
dynamics (the inverse Larmor frequency) or faster than the electron transfer
time defined by voltage. Let us define the
fluctuations of spin currents (where $\sigma=\uparrow,\downarrow$ for a chosen
quantization axis), $\Delta I^{\sigma}\left(  t\right)  =I^{\sigma}\left(
t\right)  -\left\langle I^{\sigma}\right\rangle $ and the corresponding noise
power
\begin{equation}
S^{\sigma\sigma^{\prime}}=\frac{1}{2}\left\langle \Delta I^{\sigma}\left(
t\right)  \Delta I^{\sigma^{\prime}}\left(  t^{\prime}\right)  +\Delta
I^{\sigma^{\prime}}\left(  t^{\prime}\right)  \Delta I^{\sigma}\left(
t\right)  \right\rangle .
\end{equation}
Since $I\left(  t\right)  =I^{\uparrow}\left(  t\right)  +I^{\downarrow
}\left(  t\right)  $, the charge noise power can be written $S^{\mathrm{ch}%
}=\sum\limits_{\sigma,\sigma^{\prime}=\uparrow,\downarrow}S^{\sigma
\sigma^{\prime}}.$ The spin current polarized perpendicular to $\mathbf{m}$
transfers angular momentum to the ferromagnet almost instantaneously at the
interface. Taking the spin quantization axis in the direction of
$\widehat{\theta},$ \textit{i.e.} perpendicular to the magnetization of
$F_{\mathrm{free}}$, there is therefore no correlation between the
(transverse) spin currents $I_{r}^{\sigma}\left(  t\right)  $ on both sides
($r=L,R$) of $F_{\mathrm{free}}$. The component of the field $\mathbf{h}$
relevant for our discussion is $h_{\theta}$, transverse to the magnetization,
and can be expressed as $h_{\theta}\left(  t\right)  =\sum\limits_{r=L,R}%
(\gamma\hbar/2e{\cal M}_{S})\left(  I_{r}^{\uparrow}\left(  t\right)
-I_{r}^{\downarrow}\left(  t\right)  \right)  $. 

We describe the switching process in the N\'{e}el-Brown model of thermally
assisted magnetization reversal \cite{brown63}. We consider a statistical
ensemble of particles on the unit sphere that represents the probability
density for the magnetization direction. Under the influence of the (random)
forces these particles diffuse over the sphere. A Fokker-Planck equation
describes the evolution of the probability density $P\left(  \theta
,\varphi,r\right)  $ to find the magnetization vector in a certain direction
\cite{brown63,apalkov04}.%

\begin{align}
\frac{\partial P}{\partial t} &  =-\frac{\partial}{\partial\theta}A_{\theta
}P+\frac{1}{2}\frac{\partial^{2}}{\partial\theta^{2}}B_{\theta
\theta}P\nonumber\\
&  -\frac{\partial}{\partial\varphi}A_{\varphi}P+\frac{1}{2}\frac{\partial
^{2}}{\partial\varphi^{2}}B_{\varphi\varphi}P,
\end{align}
with $A_{\theta}=-\gamma\alpha H_{K}\cos\theta\sin\theta-\gamma J_{S}%
\sin\theta+\frac{1}{2}\mu_{\theta\theta}\cot\theta,$ $B_{\theta\theta}=\mu
_{\theta\theta},\ A_{\varphi}=\gamma H_{K}\cos\theta$ and $B_{\varphi\varphi
}=\mu_{\varphi\varphi}/\sin^{2}\theta$. The diffusion constant $\mu_{\theta\theta}$
is affected by the noise power $S_{w}.$ We consider the
limit of small polarization $p$ in which $\mu_{\theta\theta}$ becomes
independent of $\theta$. The value of $\mu_{\varphi\varphi}$ is then equal to 
$\mu_{\theta\theta}$, but is not relevant for the calculation of the switching time. The new diffusion parameter can then be written in
terms of $S_{w}$: 
\begin{equation}
\mu_{\theta\theta}=\mu_{\mathrm{thermal}}+\frac{1}{2}\left(  \frac{\gamma\hbar
}{e{\cal M}_{S}}\right)  ^{2}S_{w},\label{Mnoise}%
\end{equation}
where
\begin{equation}
\mu_{\mathrm{thermal}}=\frac{2\gamma\alpha k_{B}T}{{\cal M}_{S}
}\label{muDthermal}%
\end{equation}
is the N\'{e}el-Brown diffusion constant. 

\begin{figure}[tbh]
\centering
\includegraphics[width=0.95\columnwidth]{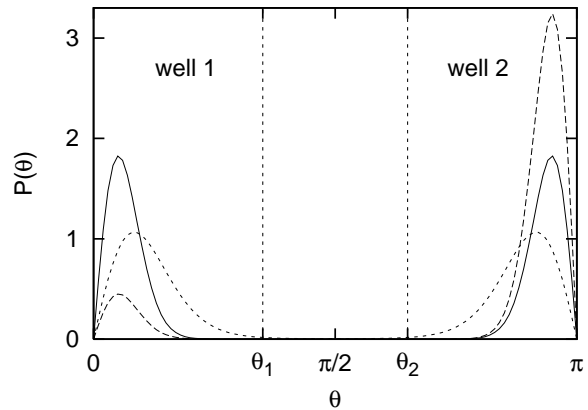} .\caption{The
probability density for the magnetization direction in the steady state as a
function of the angle $\theta$. The density in the absence of a current is
shown for a noise spectral density of $S_{w}=3.0\cdot10^{-20}C^{2}s^{-1}$
(solid) and $S_{w}=1.0\cdot10^{-19}\mathrm{C}^{2}\mathrm{s}^{-1}$ (dotted). The
asymmetry in the dashed curve is caused by a current of $0.1\;\mathrm{mA}$
with noise spectral density $S_{w}=3.0\cdot10^{-20}\mathrm{C}^{2}\mathrm{s}%
^{-1}$.}%
\label{Fokker-Planck}%
\end{figure}

We introduce the surface probability current $\mathbf{j}\left(  \theta
,\varphi\right)  $ defined as $\partial W/\partial t=-\nabla\cdot\mathbf{j}%
\ $in terms of the surface probability density $W\left(  \theta,\varphi
\right)  =P\left(  \theta,\varphi\right)  /\sin\theta$. In the steady state
\begin{equation}
\mathbf{j}\cdot\hat{\theta}=A_{\theta}W-\frac{1}{2\sin\theta}\frac{\partial
}{\partial\theta}B_{\theta\theta}W\sin\theta\equiv0.
\end{equation}
Since $W$ does not depend on $\varphi$,%

\begin{equation}
W\left(  \theta\right)  \propto w\left(  \theta\right)  =\exp\frac
{-\gamma\alpha H_{K}\sin^{2}\theta+2\gamma J_{S}\cos\theta}{\mu_{\theta\theta}}.
\end{equation}
In Fig. \ref{Fokker-Planck} the probability density is shown for some combinations of
current and noise power. It can be seen that applying a current results in an
asymmetric distribution because the spin torque drives the magnetization to
the second well, whereas an increased noise power broadens the distribution in
the wells.

\begin{figure}[tbh]
\centering
\includegraphics[width=0.9\columnwidth]{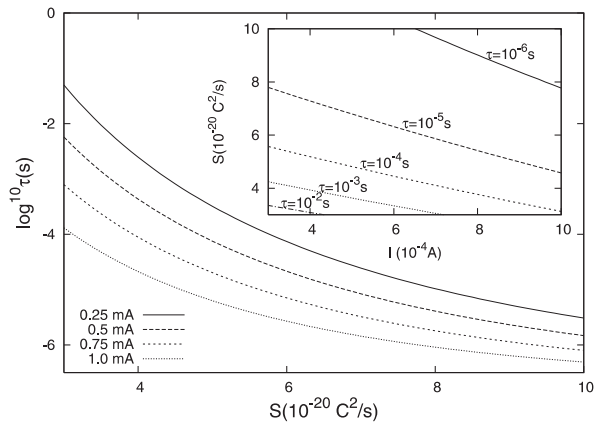}\caption{Switching
time as a function of noise spectral density for several values of the current. Inset: 
Equal-switching-time contours as a function of current and applied noise spectral density.}%
\label{switching}%
\end{figure}

For high barriers, Kramer's method can be applied to find the switching time
of the magnet. In that limit equilibrium is attained separately in the regions
$0\leq\theta\leq\theta_{1}$ and $\theta_{2}\leq\theta\leq\pi$ (potential well
$1$ and $2$)$.$ The effective potentials at $\theta_{1}$ and $\theta_{2}$
should be separated from the potential minimum by several thermal energies
$k_{B}T$ to ensure that the probability to find the magnetization in the
middle region is low. The probability current $I_{m}$ in this middle region
defined as $\mathbf{j}\cdot\hat{\theta}=I_{m}/\left(  2\pi\sin\theta\right)
,$ is then small and does not disturb the equilibrium distribution in the two
wells significantly. In the region $\theta_{1}\leq\theta\leq\theta_{2}$ (see Fig.
\ref{Fokker-Planck}) then
\begin{equation}
\dot{n}_{1}=-\dot{n}_{2}=-\frac{n_{1}}{\tau_{1}}+\frac{n_{2}}{\tau_{2}}%
=-I_{m},
\end{equation}
where $n_{\varsigma}$ is the probability to be in well $\varsigma=1,2$. The
escape time from well $1,$
\begin{equation}
\tau_{1}=\frac{2}{\mu_{\theta\theta}}\int\limits_{0}^{\theta_{1}}d\theta w\left(
\theta\right)  \sin\theta\int\limits_{\theta_{1}}^{\theta_{2}}d\xi 
\left[w\left(
\xi\right)\sin\xi\right]^{-1}\label{timeintegrals}%
\end{equation}
is a good estimate for the switching time. When $2\gamma\alpha H_{K}/\mu
_{D},\left\vert \alpha H_{K}/J_{S}\right\vert $ $\gg1$, we can calculate the
integrals using Taylor expansions up to second order in $\theta$ and
approximating $\sin\theta$ by $\theta$ (Ref. \onlinecite{brown63}):%

\begin{equation}
\tau_{1}=\sqrt{\frac{\pi\mu_{\theta\theta}}{\gamma\alpha H_{K}}}\frac{1}{\gamma\alpha
H_{K}+\gamma J_{S}}\exp  \frac{\gamma \alpha H_{K}\left(  1-J_{S}/\alpha
H_{K}\right)^{2}}{\mu_{\theta\theta}}  .\label{time}%
\end{equation}
We see that the spin-transfer torque reduces the effective barrier height
\cite{apalkov04}, while the fluctuations increase the effective temperature,
exponentially reducing the escape time. Because of the uniaxial symmetry
$\tau_{2}\rightarrow\tau_{1}$ with $J_{S}\rightarrow-J_{S}.$ When a clearly 
preferred switching direction is required, 
$\mu_{\theta\theta}$ should be kept smaller than 
$\gamma J_{S}$. But when current and noise can be switched off immediately 
after a change in the magnetoresistance signifies that switching has occured, this is no longer a constraint.
By combining Eq.
(\ref{time}), (\ref{muDthermal}) and (\ref{Mnoise}) we can establish an
analytical relation between the applied noise power, current and switching
time. We use Eq. (\ref{timeintegrals}) to evaluate the switching time 
as a function of current and noise spectral density (see Fig. \ref{switching}%
). The model is here at room temperature with parameters typical
for real spin valves. The ferromagnet $F_{\mathrm{free}}$ is specified by an
anisotropy field $H_{K}=50$ $\mathrm{mT}$ and total magnetic moment
${\cal M}_{S}=10^{-17}\mathrm{Am}^{2}.$ The interface resistances are taken to be $1$
$\mathrm{\Omega},$ the polarization $p$ of the contacts $p=0.2$, and the
damping constant $\alpha=0.02$. A typical noise generator power is
$15$ dB higher than the thermal noise power of a $50\;\mathrm{\Omega}$ resistor at
$295\;\mathrm{K}$. This gives a value of $S_{w}=1.0\cdot10^{-20}%
\;\mathrm{J/\Omega}$. Higher noise powers, corresponding to a thermal 
noise at thousands of Kelvins, can be readily generated.

From Fig. \ref{switching} we see that by keeping the current fixed and
increasing the noise, the switching time can be reduced by orders of
magnitude. The contours in the inset to the figure 
show how the current can be reduced by increasing the noise when the switching
time is kept fixed. The power dissipated by
the system is proportional to $\left\langle I\left(  t\right)  ^{2}\right\rangle
$, which increases with external noise by $S_{\omega}\Delta\omega/\pi$. Since the
typical bandwidth $\Delta\omega$ is of the order of $1$ GHz, the main
contribution to the power consumption comes from the average switching current. 
We observe that the power can be reduced by an order of magnitude. 

The calculations presented here are restricted to the high-barrier
approximation, which sets limits to the currents and effective temperatures.
Still, the mechanism of noise-assisted magnetization switching should be
useful in other regimes as well. The experimental realization of this
mechanism might face some difficulties such as enabling the high-frequency
external fluctuations to reach the system. For instance, in the case of a
large environmental capacitance between the leads, the impedance mismatch to
the device might require additional measures. Another application of 
this system would be the measurement of the noise level: by calibration 
of the device, an unambiguous relation between current, switching time and 
noise power can be established.

In conclusion, we propose an energy-efficient scheme for current-induced
magnetization switching that is assisted by noise. Our approach is based on
solving stochastic equations depending on the spectral density of an external
noise source. The solution of the corresponding Fokker-Planck equation gives
the dependence of the switching time on current and noise level. The current
necessary to switch the magnetization can be reduced by applying externally
generated current fluctuations. Without importantly complicating the device
architecture the efficiency of spin-transfer torque devices can be improved by
exponentially reduced switching times and an order-of-magnitude smaller
power consumption. This could make the difference for the attractiveness of the
current-induced switching mechanism for real-life applications.

This work is supported by the \textquotedblleft Stichting voor Fundamenteel Onderzoek der Materie\textquotedblright\ 
(FOM), and the \textquotedblleft Nederlandse Organisatie voor Wetenschappelijk Onderzoek\textquotedblright\ (NWO).
We would like to thank Ya. M. Blanter, A. Brataas and Yu. V. Nazarov for discussions.

\end{document}